\documentclass[
  aip,
  reprint,
  superscriptaddress,
  amsmath,amssymb
]{revtex4-2}

\usepackage{graphicx}
\usepackage{dcolumn}
\usepackage{bm}

\usepackage[utf8]{inputenc}
\usepackage{siunitx}
\usepackage{physics}
\usepackage{xfrac}
\usepackage{appendix}
\usepackage[usenames,dvipsnames]{xcolor}

\begin{document}

\title{Dressed-State Quantum Information Processing Using a Multilayer Ion Trap Chip with an Integrated Radio Frequency Resonator}

\author{Ivan A. Boldin*}
\affiliation{Department of Physics, School of Science and Technology, University of Siegen, 57068 Siegen, Germany}
\altaffiliation[These authors contributed equally to this work. Present address: ]{eleQtron GmbH, Heeserstr. 5, 57072 Siegen, Germany}

\author{Elham Esteki*}
\affiliation{Department of Physics, School of Science and Technology, University of Siegen, 57068 Siegen, Germany}
\altaffiliation[These authors contributed equally to this work. Present address: ]{eleQtron GmbH, Heeserstr. 5, 57072 Siegen, Germany}

\author{Bogdan Okhrimenko}
\affiliation{Department of Physics, School of Science and Technology, University of Siegen, 57068 Siegen, Germany}

\author{Friederike Giebel}
\affiliation{Physikalisch-Technische Bundesanstalt, Bundesallee 100, 38116 Braunschweig, Germany}
\affiliation{Institute of Quantum Optics, Leibniz Universit\"at Hannover, Schneiderberg 39, 30167 Hannover, Germany}

\author{Konstantin Thronberens}
\affiliation{Physikalisch-Technische Bundesanstalt, Bundesallee 100, 38116 Braunschweig, Germany}
\affiliation{Institute of Quantum Optics, Leibniz Universit\"at Hannover, Schneiderberg 39, 30167 Hannover, Germany}

\author{Amado Bautista-Salvador}

\author{Christian Ospelkaus}
\affiliation{Physikalisch-Technische Bundesanstalt, Bundesallee 100, 38116 Braunschweig, Germany}
\affiliation{Institute of Quantum Optics, Leibniz Universit\"at Hannover, Schneiderberg 39, 30167 Hannover, Germany}

\author{Christof Wunderlich}
\affiliation{Department of Physics, School of Science and Technology, University of Siegen, 57068 Siegen, Germany}

\date{\today}

\begin{abstract}

We present the experimental preparation of a radio frequency (RF)-dressed-state qubit, the coherent operation of a single RF-dressed qubit, and the selective readout of RF-dressed hyperfine states of $^{171}$Yb${^+}$ ions trapped using a novel microstructured ion trap chip. Additionally, we demonstrate the individual addressing of RF-dressed-state qubits in a two-ion Coulomb crystal. To this end, we developed, manufactured, and characterized a trap chip specifcally designed for quantum information processing with RF-dressed-state qubits. The chip consists of multiple layers, one of which includes an integrated resonator that creates a gradient of the RF magnetic field amplitude along the axis of a linear ion crystal. This RF gradient enables the individual addressing of dressed qubit states as well as their coupling. We characterize the ion trap chip experimentally using the hyperfine states of the electronic ground state of laser cooled $^{171}$Yb${^+}$ ions and by applying a single, global, modulated RF field oscillating near 12.6 GHz that drives dressed-state resonances.

\end{abstract}

\maketitle

\section{Introduction}
For storing and processing quantum information with trapped atomic ions, often the internal states of ions are used together with their motional states, with both degrees of freedom controlled by laser light \cite{Cirac1995, Monroe1995, Solano1999, Molmer2000, Milburn2000, Leibfried2003, Schmidt-Kaler2003B, Pirandola2006, Travaglione2002, Fluehmann2018}. The typical energy separation $h\nu$ (with Planck's constant $h$) of the ionic states used for quantum information processing corresponds to frequencies $\nu$ from the radio frequency (RF) regime to the optical regime. Selecting hyperfine states with resonances in the RF regime as physical states for quantum information processing is useful, because of their long natural lifetime. In addition, this frequency range is particularly suitable for coherent manipulation of quantum states, as many pieces of hardware are readily available, exhibit high frequency and phase stability, and can be miniaturized.

Transitions between hyperfine qubit states are driven directly with a resonant RF field \cite{Mintert2001, wunderlich2002conditional, Ospelkaus2008, Johanning2009, Ospelkaus2011, Khromova2012, weidt2016trapped, Harty2016, Sutherland2019, srinivas2019trapped} or, after translating RF signals into the optical regime, using a pair of Raman laser beams \cite{Monroe1995}. For implementing conditional quantum dynamics with at least two qubits when driving ionic transitions, the necessary interaction between internal qubit states requires coupling of the qubit states to the motion of ions, since motional degrees of freedom are used as a quantum bus to transfer quantum information between qubits.

When trapped ions are coherently controlled directly with RF fields, internal qubits interact with motional states via magnetic gradient induced coupling (MAGIC). There are different branches of experimentally realized MAGIC, distinguished by the time-dependence of the magnetic gradient field, which can be static \cite{Mintert2001, wunderlich2002conditional, Khromova2012, weidt2016trapped, Piltz2016}, oscillating at an RF frequency close to the qubit resonance \cite{Ospelkaus2008, Ospelkaus2011, Harty2016, wolk2017quantum, zarantonello2019robust}, or oscillating at a frequency close to a motional resonance \cite{Sutherland2019, srinivas2019trapped}.

In the case of a static magnetic field gradient, ions can be individually coherently addressed in frequency space with low crosstalk, as the qubit resonances are individually Zeemann-shifted depending on the location of an ion \cite{Johanning2009, kunert2014planar, piltz2014trapped, Randall2015, Kawai2017}. As a drawback, one needs to make efforts to isolate the system from magnetic field noise, as it directly affects the coherence time. If the magnetic field gradient oscillates near the qubit transition frequency, then magnetic field insensitive states can be used, thus reducing the qubits' sensitivity to magnetic field noise. However, the coupling mechanism requires that the RF magnetic field amplitude be near-zero at the ions' trapping positions, while the RF magnetic field amplitude gradient should be high \cite{Ospelkaus2011, warring2013techniques, carsjens2014surface} (although dynamical decoupling can be applied to reduce the effect of the non-zero field \cite{harty2016high}), which can be challenging to achieve at the same time. Another challenge is the individual addressing of single qubits in a longer ion chain, although methods have been developed for this purpose \cite{warring2013individual, srinivas2021high, Sutherland2023}.

Instead of using bare electronic levels as qubits, quantum information can be encoded into eigenstates of the qubit states that interact with a driving field, the so-called dressed states. Single-qubit manipulations, state preparation, detection, and coherence times up to 5 s have been demonstrated for trapped ion dressed-state qubits using magnetic field sensitive bare states \cite{timoney2011quantum, tan2013demonstration, webster2013simple, randall2015efficient, Baumgart2017}.

Here, we report on the design, manufacturing, and experimental characterization of a novel multi-layer ion trap chip. This trap chip allows for generating an RF gradient field along the weak-confinement axis of a Coulomb crystal. The RF field is used to create dressed qubit states with varying energy splittings along this axis. After motivating the work presented in this article in section \ref{Motivation}, we describe in section \ref{Design} in detail the design process of the trap including a novel RF resonating structure. Section \ref{Manufacture} is dedicated to techniques used to manufacture the trap chip. Measurements with trapped $^{171}$Yb${^+}$ ions reported in Section \ref{Experiments} demonstrate the functionality of this ion trap by position-dependent Rabi oscillations between bare states, in agreement with numerical simulations of the trap structure, and, in addition, Rabi oscillations between dressed qubit states in a two-ion crystal.

\section{Dressed-states qubit created by an RF field with axial gradient}
\label{Motivation}
In \cite{wolk2017quantum} it is shown that trapped-ion Coulomb crystals subject to either a static magnetic field gradient plus a spatially uniform qubit driving field, or a spatially varying resonant oscillating magnetic field are described by essentially identical Hamiltonians in an appropriate dressed-state picture. Based on this equivalence, it is proposed in \cite{wolk2017quantum} to use atomic states dressed by an oscillating magnetic field with a gradient along the axial direction of a linear Coulomb crystal to implement conditional quantum gates. This eliminates the need to null the oscillating magnetic field at the ions' positions, thus simplifying coherent control, especially with many ions. Dressed qubits are insensitive to ambient magnetic noise \cite{timoney2011quantum}, eliminating the need for strong and stable bias fields to create ``clock'' qubits robust against decoherence by electromagnetic noise.

The gradient of the RF magnetic field amplitude also leads to a coupling between the dressed state qubit state and the motion of the ions \cite{wolk2017quantum}, and thus multi-qubit gates are possible. The situation is analogous to static MAGIC \cite{Mintert2001, wunderlich2002conditional}, but the magnetic field sensitive hyperfine states are replaced with dressed states, and the static magnetic field gradient is replaced by the gradient of the RF magnetic field amplitude. Therefore, a long-range spin-spin coupling between dressed qubits could be created using a single oscillating gradient, or quantum gates employing double-dressed states could be carried out \cite{Nuennerich2025}. This in turn can be exploited for quantum simulations and quantum computation.

The Lamb-Dicke parameter $\epsilon \propto 1/\nu^{3/2}$ quantifies the coupling strength between the internal ionic states and the motional states of the ions \cite{wolk2017quantum}, where $\nu$ is the eigenfrequency of the motional normal mode under consideration. Since frequencies of axial modes are smaller than radial mode frequencies, applying the gradient along the weak-confinement axis instead of along the radial direction of a linear Coulomb crystal can significantly enhance the qubit-motion coupling $\epsilon$.

Furthermore, this method enables for straightforward individual addressing of ions with low cross-talk: Each qubit in a linear Coulomb crystal experiences a different RF amplitude and thus is driven at a different Rabi frequency, which in turn makes the energy separation between the dressed states different for each dressed qubit. Thus, dressed qubits can be individually addressed by choosing the appropriate frequency of an additional RF field driving transitions between dressed states. Alternatively, the RF field creating the dressing can be modulated at the desired dressed qubit resonance frequency to control the dressed state qubit.

\section{Trap Chip for creating an axial oscillating gradient field}

\subsection{Design of the trap chip}
\label{Design}
A crucial ingredient to implement the dressed states scheme as described above is a gradient of the RF magnetic field amplitude along the trapping axis of the ion trap. For achieving this, we decided on a multi-layer ion trap chip design. The bottom layer (L1) of the trap chip contains the RF resonator for the qubit driving field and DC electrodes, while the top layer (L2) contains the RF$_{\text{trap}}$ electrodes needed to create the effective trapping potential. The metal layers are separated by a dielectric layer (V1) that contains the vertical interconnect access (VIAS) between L1 and L2.

\begin{figure}
  \includegraphics[width=\columnwidth]{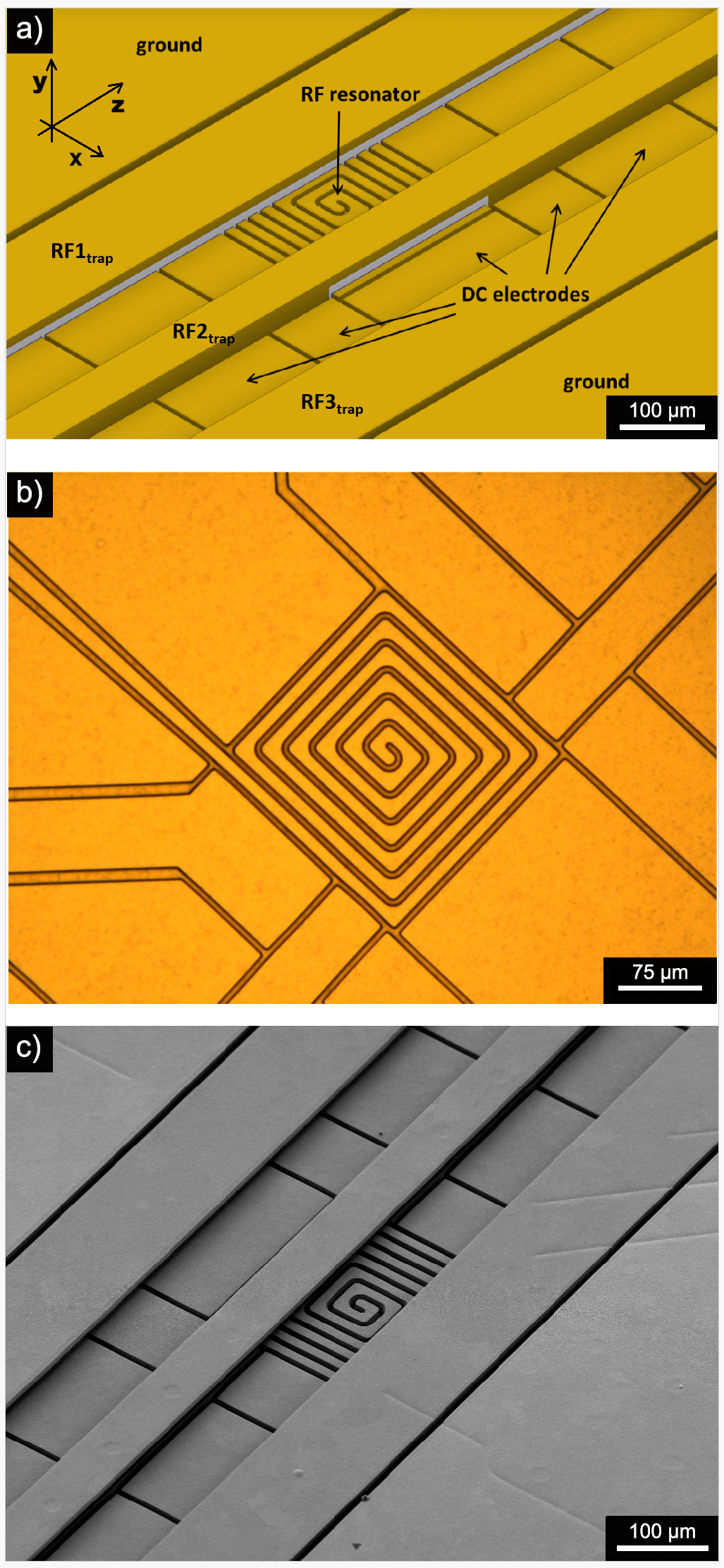}
  \caption{\label{fig:sem_pic} a) The central part of the ion trap chip (CAD model). The RF resonator is seen in the bottom layer, the three electrodes for generating the radial electric trapping field are labeled RF1$_{\text{trap}}$, RF2$_{\text{trap}}$ and RF3$_{\text{trap}}$. Not all of the DC electrodes used for axial confinement of the ions are indicated with arrows. b) An optical microscopy picture of the bottom layer of the ion trap chip. The picture was taken during the production process, before the other layer was produced on top. c) A SEM micrograph from the center of the ion trap chip, belonging to the same batch as the one use for the experiments reported here. The upper layer, L2 is aligned with a precision below 0.5~$\mu$m with respect to the bottom layer, L1 in the x-direction while in the y-direction the alignment precision is around 1~$\mu$m.}
\end{figure}

The RF electrode is a spiral, which is open at its inner end and is connected to an RF feedline at its other end (Fig.~\ref{fig:sem_pic}). At a point where the spiral transitions into the feed-line, the RF power leaving the spiral is partially reflected back into it. For a particular frequency, the reflected wave matches in phase with the one coming from the feedline as well as with the further reflections, amplifying the current in the spiral and making it a resonator. The spiral works as a coil to maximize the amplitude of the RF magnetic field at a point above its center and, at the same time, to maximize the gradient of amplitude of the RF magnetic field in a region above the edge of the spiral.

The RF electrode is designed so that it resonates at a particular frequency that can be chosen to match the frequency of the qubit transition. The resonance is achieved when the length of the coil is around a quarter of the wavelength. A simulation-based analysis was performed to choose the coil dimensions values that result in the desired resonance frequency.

The resonator is designed to drive the hyperfine transition in $^{171}$Yb${^+}$ ions at a frequency near 12.64~GHz, corresponding to a wavelength of 2.4~cm in vacuum and around 1~cm on the ion trap chip. RF simulations were performed on Ansys HFSS. The spiral geometry is defined by the width of the spiral line, the gap between the turns, and the number of turns. Using simulations, these three parameters were optimized to maximize the amplitude of the RF magnetic field at the trapping position. In order to make the fabrication easier, we decided to limit the gap to be not less than 5~$\mu$m and the width of the spiral line to be not less than 10~$\mu$m. The thickness of the L1 layer was chosen to be 10~$\mu$m. The RF magnetic field amplitude grows with the number of turns; however, as the length of the coil should stay constant, the width of the spiral line or the gaps between the turns have to get smaller. Taking into account the limitations mentioned above, the optimization resulted in a spiral line width of 11.8~$\mu$m, a gap of 5~$\mu$m between turns, and the spiral making 6 turns. These parameters were chosen for the final design.

The upper layer of the trap (L2) contains three RF$_{\text{trap}}$ electrodes to create the effective axial potential for ion trapping (Fig.~\ref{fig:sem_pic}a). Although two RF$_{\text{trap}}$ electrodes would be sufficient for ion trapping, we have chosen to include three electrodes with individually tunable RF$_{\text{trap}}$ voltages. Such a configuration offers the following options: i) with RF applied to electrodes RF1$_{\text{trap}}$ and RF2$_{\text{trap}}$, the system becomes an asymmetric ion trap for trapping at a lower trapping height, namely 52~$\mu$m above the upper layer of the chip (defined as position A); ii) with RF applied to electrodes RF1$_{\text{trap}}$ and RF3$_{\text{trap}}$, trapping at a higher height is possible (149~$\mu$m above the upper layer of the chip (position B); iii) by applying RF of some amplitude to RF1$_{\text{trap}}$ and RF3$_{\text{trap}}$ and an RF of a different amplitude to RF2$_{\text{trap}}$ one can trap two parallel strings of ions and vary the separation between them \cite{tanaka2014design}. Only option i) was used in this work.

The widths of electrodes RF1$_{\text{trap}}$ and RF3$_{\text{trap}}$ are 117.5~$\mu$m, while the width of electrode RF2$_{\text{trap}}$ is 55~$\mu$m. The width of the opening between the electrodes RF1$_{\text{trap}}$ and RF2$_{\text{trap}}$ is 95~$\mu$m. All gaps between electrodes are 5~$\mu$m wide, both in the top layer and in the bottom layer. Between the RF$_{\text{trap}}$ electrodes, the top layer is completely removed to expose the ions to the RF field of the coil and the DC field from the DC electrodes. An advantage of such a design is that at position A there is no scattered light from the region of the trap right below the ions because this part is in the shadow.

\begin{figure}
  \includegraphics[width=\columnwidth]{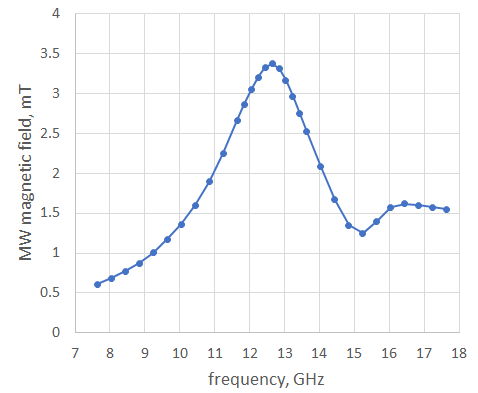}
  \caption{\label{fig:resonance_curve} Simulated RF magnetic field amplitude at position A (72~$\mu$m above the spiral center) for 1~W of RF power applied to the chip, depending on the frequency.}
\end{figure}

The simulated RF magnetic field amplitude along the trapping axis (position A) depending on the RF frequency is shown in Fig.~\ref{fig:resonance_curve}. The Q-factor of the resonator derived from the full width at half maximum of this curve is 5.

\subsection{Manufacturing the trap chip}
\label{Manufacture}
The ion trap chip is manufactured using a multi-level metalization process, as described in \cite{bautista2019multilayer}. The fabrication procedure was adapted to an aluminum-nitride (AlN) substrate, chosen for better heat dissipation. The method starts by thermally depositing a 17-\SI{}{\nano\m}-thin layer of Ti and a 50-\SI{}{\nano\m}-thin layer of Au on top of an AlN substrate (180 \SI{}{\watt\per\meter\kelvin}). The thin Ti layer enhances the adhesion of subsequent layers, whereas the thin Au film works as a starting conductive layer for a subsequent electroplating step. After coating the AlN substrate on top of Au/Ti with a negative resist, we performed UV photo-lithography. In general, this allows structures with critical dimension (CD) $>$ 0.5~$\mu$m to be resolved. Due to the polycrystalline nature of AlN substrates, CD values on this type of substrate are approximately 1~$\mu$m. After the coated substrate is exposed to UV light, the resist is developed. The resulting patterned resist mold is immersed in a cyanide-free gold solution and a lower interconnect level L1 is electrochemically grown up to 10~\SI{}{\micro\meter}. After resist stripping, a subsequent UV photolithography is performed on top of L1. An electroplating process is used to define the intermediate via layer V1. After these two electroplating steps, a global dry etching is performed to remove the starting layer (Au) and adhesion layer (Ti) by Ar bombardment. On top of the now electrically isolated electrodes, we spin-coat a thick film of polyimide (PI) 2000 series.

The topography of the first metallic layer L1 is transferred to the PI, resulting in poor flatness. To reset the planarity of the upper layer on top of the PI to a flatness $\leq$ 1~$\mu$m, we perform a chemical-mechanical process (CMP) on top of the PI (20~$\mu$m thickness). Before growing the L2 layer, we perform a local etch-back combined with an additional UV photolithography step to completely free the upper areas on top of the VIAS in V1. The uppermost layer L2 is now grown as previously, using UV photo-lithography and electroplating after depositing a bi-layer film of Au/Ti. Finally, after the Au/Ti seed layer was removed using dry etching, we removed the PI from the gaps in the Au layer by a fluorine-based plasma etching. This is done to minimize the exposure of the ions to dielectric surfaces, where an uncontrolled accumulation of charges can take place.

On chips resulting from the same batch, we have performed pre-assembly characterization such as SEM inspection and surface flashover tests. An SEM micrograph (Fig.~\ref{fig:sem_pic}c) shows the innermost part of the ion trap chip and a close-up view of the RF resonator with the spiral shape defined close to the ion trapping position. On the other hand, we performed surface flashover tests on similar ion trap chips. We applied a DC voltage in between all neighboring trap electrodes, ramping the DC amplitude from 0 to up to 250~V in a time lapse of five minutes, without observing any failure inspected through a CMOS camera.

\section{Experimental procedures and results}
\label{Experiments}
\subsection{Experimental setup}
\begin{figure}
  \includegraphics[width=\columnwidth]{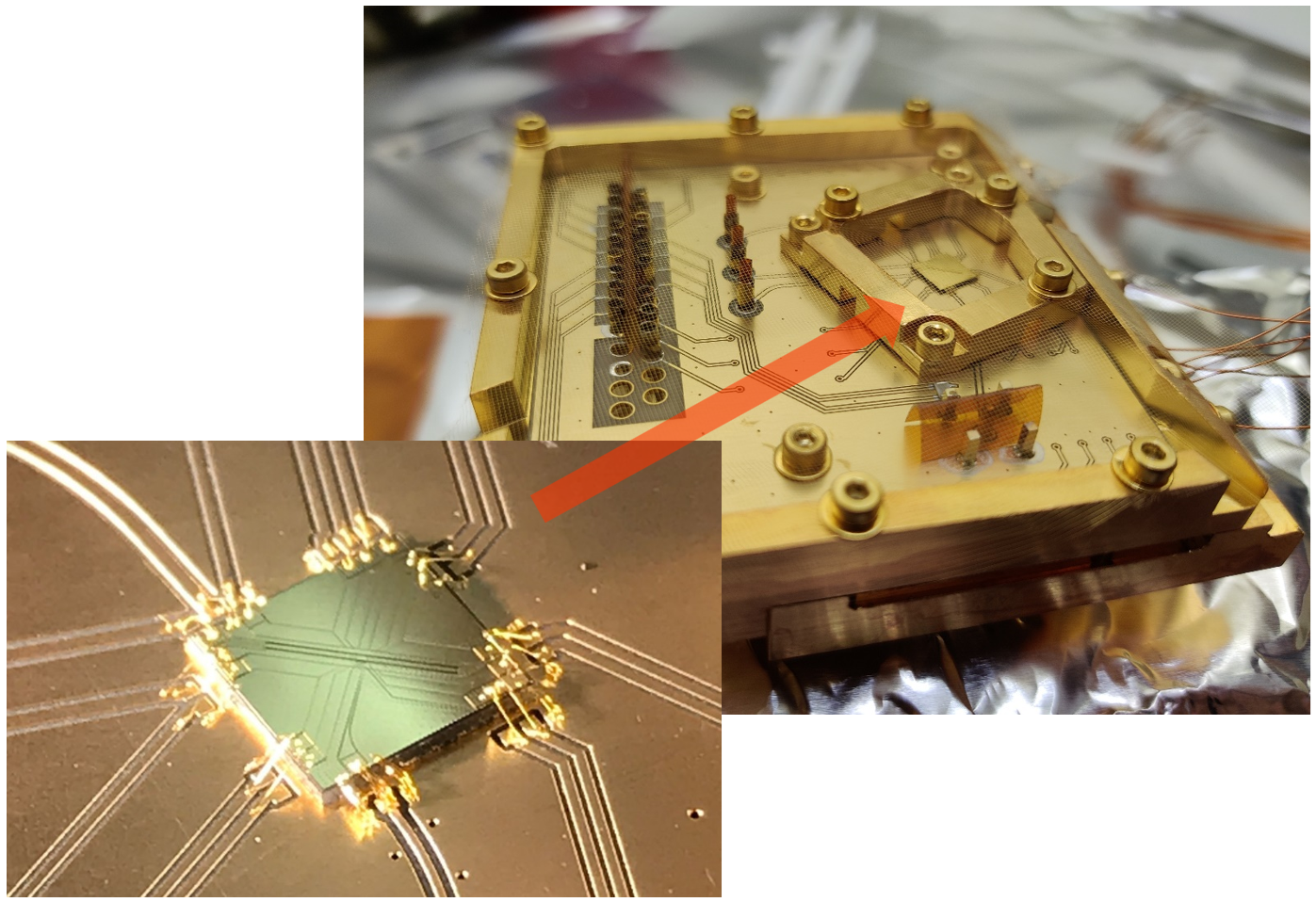}
  \caption{\label{fig:assembly_pic} Ion trap chip with RF resonator at its center mounted on the chip-carrier board.}
\end{figure}

The trap chip is a square with 5~mm sides. It is mounted on a custom-made aluminum nitride chip-carrier board. The chip is glued to the board and electrically connected by wire-bonding. On the board, there are RC-filters for each of the DC electrodes with a cut-off frequency of 1~kHz. There are two Faraday cages on the chip carrier: one encloses the whole chip carrier board, while the other only surrounds the chip, but not the elements of the RC-filters. The Faraday cages are completed by a gold mesh (90\% transmission) covering both of them from the top (Fig.~\ref{fig:assembly_pic}).

We typically apply an amplitude of 25~V at a frequency of 15~MHz to electrodes RF1$_{\text{trap}}$ and RF2$_{\text{trap}}$. The ions' radial secular frequency at this voltage amplitude is around 1~MHz and the axial secular frequency is typically in the range 200--300~kHz.

The RF signal driving atomic states is supplied through an SMA connector mounted on the chip-carrier board, which is connected to a 50~Ohm feed-line, which also has an impedance matching segment right before the chip. The matching segment is a segment of the RF feed-line with a different width and gap size, and thus a different characteristic impedance. RF simulations of the entire chip and chip-carrier board system were carried out using Ansys HFSS. The width and gap size of the RF feed line and the coupling segment, as well as the length of the coupling segment, were optimized on the basis of these simulations, with an objective of maximizing the RF magnetic field amplitude above the spiral for a given RF power applied to the board.

Three lasers with wavelengths near 369~nm, 399~nm, and 935~nm, respectively, are used to trap and cool Ytterbium ions. The 369~nm laser is used for Doppler cooling, state initialization, and -detection. The 935~nm laser depletes a metastable state to which the ion occasionally decays. The 399~nm laser, together with the 369 laser, is used for two-photon ionization of Yb atoms. The details of the level structure and the transitions driven by the lasers can be found in previous work, e.g. \cite{Wunderlich2003,Johanning2011}.

The directions of the laser beams, the imaging system, including the objective and an EMCCD camera, as well as the source of ytterbium atoms are described in \cite{kunert2014planar}. The lasers are directed at a 45 degree angle to the trapping axis. Ytterbium atoms are supplied by a miniature evaporation oven located around 4~cm from the trapping region. The stream of atoms is also directed at 45 degrees to the trap axis and 90 degrees to the lasers. The fluorescence of the ions is collected with an objective with numerical aperture NA=0.4.

\subsection{RF field amplitude as a function of axial position}
Two substates of the hyperfine ground state manifold of $^{171}$Yb${^+}$ are used as bare qubit states. In the presence of an external static magnetic field $\vec{B}$, the hyperfine resonance near 12.64~GHz is split into three transitions from state $^2S_{1/2}\ket{F=0}$ to state $^2S_{1/2}\ket{F=1, m_F=-1,0,1}$. We label these resonances as $\sigma_-$, $\pi$ and $\sigma_+$, respectively, considering the angle $\alpha$ between $\vec{B}$, determining the quantization axis, and the magnetic field of the RF field. The Rabi frequencies of these three transitions for a given applied RF power depend on $\alpha$.

In order to characterize the performance of the RF resonator, we perform a series of Rabi frequency measurements on Doppler-cooled $^{171}$Yb${^+}$ ions trapped at several positions along the trapping axis (height position A). For that purpose, the qubit is initialized in state $^2S_{1/2}\ket{F=0}$ using optical pumping near 369~nm \cite{nop2021ytterbium}. Then a variable duration RF pulse, resonant with one out of three bare state qubit transitions, is applied to the RF electrode. Finally, the qubit is read out by collecting resonance fluorescence near 369~nm: if the qubit is in the $^2S_{1/2}\ket{F=1}$ state, resonance fluorescence is observed (bright state). The ion ideally does not scatter light, if the qubit is in the $^2S_{1/2}\ket{F=0}$ state (dark state) \cite{nop2021ytterbium}. The double threshold method was used to differentiate between bright and dark states; details of the detection process are given in \cite{wolk2015state}.

In order to experimentally determine the magnetic field amplitude of the RF, we have measured all three Rabi frequencies on the $^{171}$Yb${^+}$ hyperfine 12.64~GHz transition, $\Omega_-$, $\Omega_{\pi}$ and $\Omega_+$. These measurements were made at the trapping height position A at various positions of ions along the trapping axis. Then the value of $\Omega_{inv}=\sqrt{\Omega_{-}^2 + \Omega_{\pi}^2 + \Omega_{+}^2}$ was calculated for each of the measurement points. This value is invariant with respect to the static magnetic field orientation, and the RF magnetic field amplitude was found to be $\frac{\hbar}{\mu_B}\sqrt{\Omega_{-}^2 + \Omega_{\pi}^2 + \Omega_{+}^2}$ (Appendix \ref{AppB}). In this experiment, $\vec{B}$ is created by a permanent magnet and its magnitude is about 300~$\mu$T.

According to numerical simulations, 1~W of RF applied to the chip-carrier board yields RF magnetic field amplitudes at height positions A and B of 2.027~mT and 0.408~mT, respectively. This corresponds to $\Omega_{inv}$ of 28.4~MHz and 5.71~MHz on the $^{171}$Yb${^+}$ hyperfine 12.64~GHz transition.

\begin{figure}
  \includegraphics[width=\columnwidth]{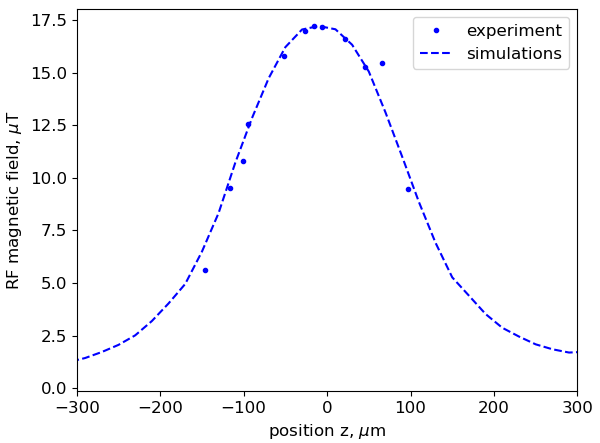}
  \caption{\label{fig:MW_profile} Microwave magnetic field amplitude calculated from Rabi frequency measurements (dots) and from the simulations (dashed line), depending on the ion position along the trapping axis (z-axis). The zero position is above the spiral center. 0.1~mW RF power was applied to the entrance to the vacuum chamber. The simulated curve was amplitude-scaled to give the best fit to the data in order to account for possible losses in the RF line; from this scaling the losses were determined to be 7.4~dB. The statistical error bars on the RF magnetic field amplitude are too small to be visible in the plot. The deviations from the simulations are likely caused by the ion-surface separation not being exactly constant for all data points.}
\end{figure}

From measurements of Rabi frequencies as a function of the axial position of an ion with 0.1~mW of RF power applied to the system (measured at the entrance to the vacuum chamber) the RF magnetic field amplitude at a given axial position is deduced. These measurement results are presented in Fig.~\ref{fig:MW_profile}. The simulated curve was amplitude-scaled to give the best fit to the data in order to account for the losses in the RF cables and connections between the vacuum feed-through and the chip-carrier. As determined by the fit, these losses sum up to 7.4~dB.

We have applied up to 10~mW of RF power and measured the Rabi frequency of 1.2300(27)~MHz at the $\pi$ transition. The chip has been designed to be able to dissipate a few Watts of applied RF power, so the maximum Rabi frequency can reach tens of MHz. As the simulations have been experimentally proven to fit the experiment well and the losses are low, one can conclude that at, for example, 4~W of applied RF power, the RF magnetic field gradient is expected to reach 25~T/m.

\subsection{Dressed state Rabi oscillations}
For convenience we recall in Appendix \ref{AppA} how the dressed states $\ket{D_-}$ and $\ket{D_+}$ of a two-level system depend on the detuning, $\Delta$ between the driving field's RF frequency and the bare state qubit resonance frequency, and on the Rabi frequency $\Omega$. A system prepared, for example, in state $\ket{0}$, during an adiabatic ramp of the driving field, will remain in $\ket{D_-}$. Similarly, the system in some superposition of the dressed states, after the driving field is adiabatically turned off, will end up in a superposition of the bare states with the same coefficients. This property can be used for the state detection.

For driving the single-qubit transitions between the dressed states, one can directly apply an oscillating magnetic field on the frequency, corresponding to the dressed states' energy separation \cite{timoney2011quantum, webster2013simple, randall2015efficient}. In this work we use a different method: We amplitude-modulate the driving field at the frequency of the dressed state qubit resonance. One can show (Appendix \ref{AppA}) that such a field drives a transition between the dressed states with Rabi frequency
\begin{equation}
  \Omega_D = \frac{m}{2}\Omega\cos(2\Theta) = \frac{m}{2}\frac{\Omega\Delta}{\Omega_g},
  \label{eq:Dressed_state_Rabi_Freq}
\end{equation}
where $m$ is the modulation depth of the RF magnetic field amplitude.

In what follows, we describe Rabi oscillations between dressed states employing a two-ion chain. Here we demonstrate, in addition, individual addressing of two dressed state qubits. The ions are placed in the region of the maximum RF magnetic field amplitude gradient, i.e. position -100~$\mu$m in Fig.~\ref{fig:MW_profile}. The separation between the two ions is 10~$\mu$m. As the initial bare qubit, we use the $\pi$-transition of the $^{171}$Yb${^+}$ hyperfine manifold.

At first, the qubits are initialized in state $^2S_{1/2}\ket{F=0}$ as in the Rabi experiment with bare states. Then we adiabatically raise the near-resonant RF field amplitude (the RF field is detuned from the $\pi$-transition by 50~kHz), the ramp taking 70~$\mu$s. As a result, each of the qubits transitions to the dressed state $\ket{D_-}$. At the end of the ramp, when the RF magnetic field amplitude reaches its maximum, the Rabi frequencies $\Omega^{(1)}$ and $\Omega^{(2)}$ of the two ions are 155.5~kHz and 200.8~kHz, respectively. After that, the modulation of the RF field amplitude at frequency $\Omega_g^{(1)} = 163.3$~kHz of the first ion is switched on and left on for a variable duration of time.

We modulate the RF amplitude with square pulses instead of a sine function to save the memory of the frequency generator. Only the first component of the Fourier series will be in resonance with the dressed-states transition. The amplitude of this component for a square wave of amplitude $A$ is $4A/\pi$, therefore, the Rabi frequency in Eq.~(\ref{eq:Dressed_state_Rabi_Freq}) should be multiplied by a factor of $4/\pi$.

In this experiment, the modulation depth is 0.1, that is, the amplitude of the driving RF field changes between 0.9 and 1.1 times the non-modulated value. Finally, the RF field is adiabatically turned off to transfer the dressed-state populations into the populations of the bare state qubits. Then the bare state qubits' populations are read out, as already described above.

\begin{figure}
  \begin{tabular}{c}
    \includegraphics[width=\columnwidth]{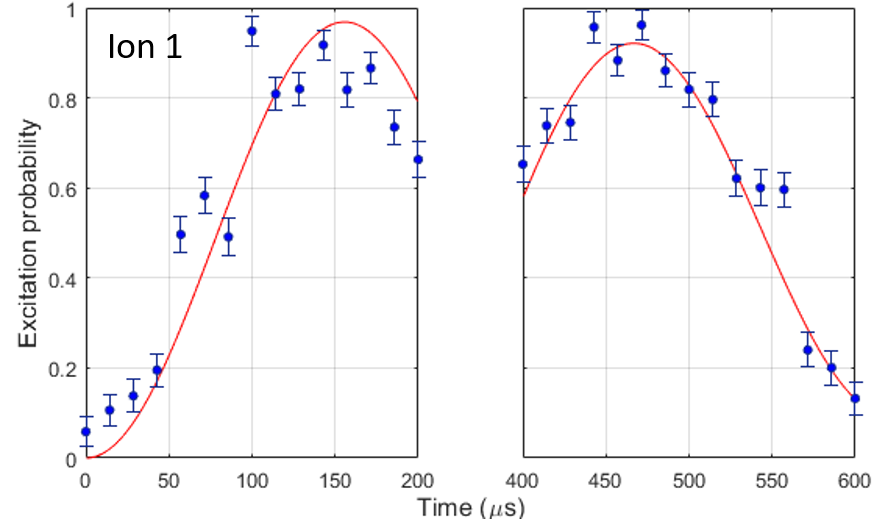} \\
    \includegraphics[width=\columnwidth]{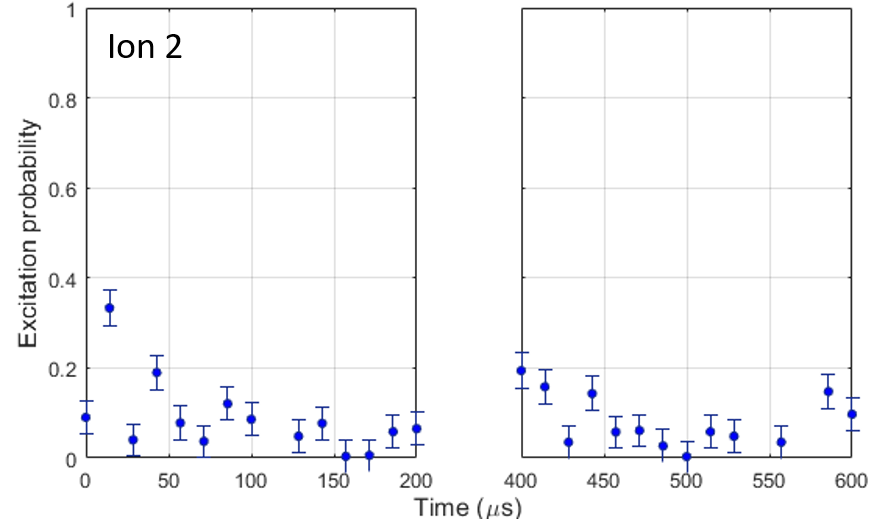} \\
  \end{tabular}
  \caption{\label{fig:Rabi} Rabi oscillations between the dressed-states of two ions. The ions are prepared in states $\ket{D_-}$, then an amplitude modulated RF field drives the Rabi oscillations. Finally, the dressed states are transferred into the bare states for the state detection, so dressed state $\ket{D_-}$ corresponds to excitation 0 and dressed state $\ket{D_+}$ corresponds to excitation 1. The RF driving field modulation frequency is adjusted to be equal to $\Omega_g$ of the first ion (top). The second ion remains not excited (bottom) as its $\Omega_g$ differs by 43~kHz from the modulation frequency. The red curve is a fit with a sine function. Note that the interval from 200~$\mu$s to 400~$\mu$s was not measured.}
\end{figure}

The resulting Rabi oscillations are shown in Fig.~\ref{fig:Rabi}. For the first ion, for which the modulation frequency is in resonance with $\Omega_g$, the Rabi frequency of the transition between the dressed states is $3.165(24)$~kHz, obtained by fitting the data in the upper graph in Fig.~\ref{fig:Rabi}. For the second ion, according to the simulations, the maximum excitation should be 0.4\% and it cannot be distinguished in this experiment Fig.~\ref{fig:Rabi} (bottom).

\section{Conclusion}
We have designed, manufactured, and experimentally characterized a novel multi-layer planar ion trap chip, featuring an RF resonator that generates an RF magnetic field amplitude gradient along the trapping axis. This chip enables dressed-state-based quantum information storage and processing, with the RF gradient being used for individual addressing and multi-qubit coupling. The resonator functions as designed, delivering a 1.2300(27)~MHz Rabi frequency on the $^{171}$Yb${^+}$ hyperfine transition with only 10~mW of applied RF power. We have shown essential single-qubit operations on dressed-state qubits, namely state preparation, single-qubit manipulation, and state detection. In addition, we have demonstrated individual addressing of dressed state qubits for a Coulomb crystal of two ions.

Future work will take advantage of this approach to trapped ion quantum computing by implementing multi-qubit quantum gates employing an oscillating RF field with axial gradient. Multiple interacting dressed qubits that are protected against decoherence will be controlled by a single RF control electrode. Thus, a novel route towards up-scaling of quantum information processing with trapped ions will be explored.

\begin{acknowledgments}
The authors acknowledge financial support from the Microwave Quantum Computation (MicroQC) project, grant \#820314. The authors thank Dorna Niroomand for useful comments on the manuscript. The authors thank for support from the clean-room staff, T. Weimann, P. Hinze and O. Kerker.
\end{acknowledgments}

\appendix

\section{Magnetic field amplitude}
\label{AppB}
For an $^{171}$Yb${^+}$ ion in a static magnetic field and an RF oscillating magnetic field, the transitions between the $^2S_{1/2}\ket{F=0}$ level and the $^2S_{1/2}\ket{F=1, m_F=-1,0,1}$ have the following Rabi frequencies: \cite{khromova2012quantum}:

\begin{subequations}
  \begin{align}
  \Omega_-  &= -\sqrt{2}\frac{\mu_B}{\hbar} B_{-} e^{-i\phi_{-}} \\
  \Omega_{\pi} &=  \frac{\mu_B}{\hbar} B_{\pi} e^{-i\phi_{\pi}}\\
  \Omega_+ &= \sqrt{2}\frac{\mu_B}{\hbar} B_{+} e^{-i\phi_{+}}\label{eq:rabi_plus},
  \end{align}
\end{subequations}

where $\mu_B$ is the Bohr magneton and

\begin{subequations}
  \begin{align}
  B_{-} e^{-i\phi_{-}} &=  \frac{1}{2}(B_x e^{-i\phi_x}+iB_y e^{-i\phi_y})\\
  B_{\pi} e^{-i\phi_{\pi}} &=  B_z e^{-i\phi_z}\\
  B_{+} e^{-i\phi_{+}} &= \frac{1}{2}(B_x e^{-i\phi_x}-iB_y e^{-i\phi_y})\label{eq:rabi_plus_sh},
  \end{align}
\end{subequations}

where $B_{\pi}$, $B_{-}$ and $B_{+}$ are the real-valued amplitudes of the RF magnetic field components parallel to the static magnetic field, that defines the quantization axis, and perpendicular, respectively. $\phi_{\pi}$, $\phi_{-}$ and $\phi_{+}$ are the correspondent phases of these components.

In our system, according to the simulations, the RF magnetic field polarization is linear, therefore phases $\phi_x$, $\phi_y$, $\phi_z$ can be set to zero. Then one can express the RF magnetic field components through the Rabi frequencies:

\begin{subequations}
  \begin{align}
  B_{x}  &=  \frac{\hbar}{\sqrt{2} \mu_B} (\Omega_+ - \Omega_-)\\
  B_{y}  &=  i\frac{\hbar}{\sqrt{2} \mu_B} (\Omega_+ + \Omega_-)\\
  B_{z}  &=  \frac{\hbar}{\mu_B} \Omega_\pi \label{eq:Bz}
  \end{align}
\end{subequations}

The amplitude of the RF magnetic field can be obtained as:

\begin{eqnarray}
  B_{RF}^2 = B_x {B_x}^\ast + B_y {B_y}^\ast + B_z {B_z}^\ast,\\
  B_{RF} = \frac{\hbar}{\mu_B}\sqrt{{\mid\Omega_+\mid}^2 + {\mid\Omega_{\pi}\mid}^2 + {\mid\Omega_-\mid}^2}
\end{eqnarray}

\section{Dressed state excitation}
\label{AppA}
In the simplest realization of the dressed-states qubit, a bare state transition is driven near its resonance. In the rotating frame co-rotating with the driving field, after rotating wave approximation, this two-level system is described by a time-independent Hamiltonian \cite{wolk2017quantum}:
\begin{equation}
H_0 = \frac{\hbar}{2}
\begin{pmatrix}
  -\Delta & \Omega \\
  \Omega & \Delta
\end{pmatrix},
\label{eq:one}
\end{equation}
where $\Delta$ is the detuning between the RF field frequency and the qubit transition frequency and $\Omega$ is the Rabi frequency of the transition. This Hamiltonian has two eigenvalues: $\pm \frac{\Omega_g}{2}$, where $\Omega_g = \sqrt{\Omega^2 + \Delta^2}$. The following eigenvectors correspond to these eigenstates:

\begin{subequations}
\begin{equation}
  \ket{D_-} = \frac{1}{\sqrt{2\Omega_g(\Omega_g+\Delta)}}
  \begin{pmatrix}
    -(\Omega_g + \Delta) \\
    \Omega
  \end{pmatrix},
\end{equation}
\begin{equation}
  \ket{D_+} = \frac{1}{\sqrt{2\Omega_g(\Omega_g-\Delta)}}
  \begin{pmatrix}
    \Omega_g - \Delta \\
    \Omega
  \end{pmatrix},
\end{equation}
\end{subequations}
$\ket{D_-}$ corresponding to the negative eigenvalue and $\ket{D_+}$ to the positive one. They can also be expressed as
\begin{eqnarray}
  \ket{D_-} =
  \begin{pmatrix}
    \sin(\Theta) \\
    \cos(\Theta)
  \end{pmatrix},
  \ket{D_+} =
  \begin{pmatrix}
    \cos(\Theta) \\
    -\sin(\Theta)
  \end{pmatrix},
\label{D_via_Theta}
\end{eqnarray}
where $\tan(2\Theta) = \frac{\Omega}{\Delta}$.

Dressed states based on a two-level system can be prepared by initializing the qubit in one of its bare states and ramping the driving field adiabatically. One can easily see that if $\Delta \neq 0$, for $\Omega \rightarrow 0$: $\ket{D_-} \rightarrow \ket{0}$, $\ket{D_+} \rightarrow \ket{1}$.

If the driving field is amplitude-modulated, the Hamiltonian from Eq.~(\ref{eq:one}) is replaced by

\begin{equation}
  H_m = \frac{\hbar}{2}
  \begin{pmatrix}
    -\Delta & \Omega(1 + m \cos(\omega_m t)) \\
    \Omega(1 + m \cos(\omega_m t)) & \Delta
  \end{pmatrix},
\end{equation}
where $m$ is the modulation depth and $\omega_m$ is the modulation frequency. This Hamiltonian can be expressed as
\begin{equation}
  H_m = H_0 + V,
  \label{Hmasasum}
\end{equation}
with
\begin{equation}
  V = \frac{1}{2} \hbar \sigma_x m \Omega \cos(\omega_m t),
\end{equation}
where $\sigma_x$ is the Pauli matrix. To transform $H_m$ into the interaction picture with $V$ as interaction, i.e. into the dressed-states basis, one has to calculate matrix elements of $V$ in this basis. Using Eq.~(\ref{D_via_Theta}), one can easily calculate them:
\begin{subequations}
  \begin{equation}
    \bra{D_-} V \ket{D_-} = \frac{1}{2} \hbar m \Omega \cos(\omega_m t) \sin(2 \Theta)
  \end{equation}
  \begin{equation}
    \bra{D_-} V \ket{D_+} = \frac{1}{2} \hbar m \Omega \cos(\omega_m t) \cos(2 \Theta)
  \end{equation}
  \begin{equation}
    \bra{D_+} V \ket{D_-} = \frac{1}{2} \hbar m \Omega \cos(\omega_m t) \cos(2 \Theta)
  \end{equation}
  \begin{equation}
    \bra{D_+} V \ket{D_+} = -\frac{1}{2} \hbar m \Omega \cos(\omega_m t) \sin(2 \Theta)
  \end{equation}
  \label{V_matrix_elements}
\end{subequations}

In order to obtain the system time-evolution in the dressed-states picture, one can write an arbitrary state of the system as
\begin{equation}
  \ket{\psi} = c_1(t)\ket{D_-} e^{-i\frac{\Omega_g}{2}t} + c_2(t)\ket{D_+} e^{i\frac{\Omega_g}{2}t}
\end{equation}
Substituting this into the Schr\"odinger equation and making use of the matrix elements (Eq.~\ref{V_matrix_elements}) one obtains
\begin{subequations}
  \begin{equation}
    \begin{aligned}
      \dot{c}_1=-\frac{i}{2}\sin(2\Theta) m\Omega\cos(\omega_m t) c_1 -\\
      - \frac{i}{2}\cos(2\Theta) m\Omega\cos(\omega_m t) e^{-i \Omega_g t} c_2
    \end{aligned}
  \end{equation}
  \begin{equation}
    \begin{aligned}
      \dot{c}_2=\frac{i}{2}\sin(2\Theta) m\Omega\cos(\omega_m t) c_2 - \\
      - \frac{i}{2}\cos(2\Theta) m\Omega\cos(\omega_m t) e^{i \Omega_g t} c_1
    \end{aligned}
  \end{equation}
\end{subequations}
As in the rotating wave approximation (for simplicity we consider the resonance case $\omega_m=\Omega_g$), one can drop fast-oscillating terms and get the Rabi frequency of the transition between the dressed states:
\begin{equation}
  \Omega_D = \frac{m}{2}\Omega\cos(2\Theta) = \frac{m}{2}\frac{\Omega\Delta}{\Omega_g}
\end{equation}

\bibliography{Chip_dressed_states}

\end{document}